*Taylor and Francis Book Chapter*

*LOC Page*

# Contents

# Communication Traffic Characteristics Reveal an IoT Device's Identity

**Rajarshi Roy Chowdhury**

*Metropolitan University, Department of Computer Science and Engineering, Bateshwar, Sylhet-3104, Bangladesh.*

**Debashish Roy**

*Toronto Metropolitan University, School of Information Technology Management, Toronto, Canada.*

**Pg Emeroylariffion Abas**

*Universiti Brunei Darussalam, Faculty of Integrated Technologies, Jalan Tungku Link, Gadong BE1410, Brunei Darussalam.*

CONTENTS

Internet of Things (IoT) is one of the technological advancements of the twenty-first century, which can improve living standards. However, it also imposes new types of security challenges, including device authentication, traffic types classification, and malicious traffic identification, in the network domain. Traditionally, internet protocol (IP) and media access control (MAC) addresses are utilized for identifying network-connected devices in a network, whilst these addressing schemes are prone to be compromised, including spoofing attacks and MAC randomization. Therefore, device identification using only explicit identifiers is a challenging task. Accurate device identification plays a key role in securing a network. In this paper, a supervised machine learning-based device fingerprinting (DFP) model has been proposed for identifying network-connected IoT devices using only communication traffic characteristics (or implicit identifiers). A single transmission control protocol/internet protocol (TCP/IP) packet header features have been utilized for generating unique fingerprints, with the fingerprints represented as a vector of 22 features. Experimental results have shown that the proposed DFP method achieves over 98% in classifying individual IoT devices using the UNSW dataset with 22 smart-home IoT devices. This signifies that the proposed approach is invaluable to network operators in making their networks more secure.

## 1.1 INTRODUCTION

In the era of the Internet of Things (IoT), IoT has brought new opportunities and, with them, new forms of security and privacy threats in cyberspace. Traditional approaches and methods of securing a network are insufficient to counter these new forms of threats. IoT devices are commonly designed and built for specific purposes and, as such, are commonly resource-constrained in terms of processing power, memory, and energy capacities [1, 2, 3, 4]. This is in contrast to general-purpose computing devices, referred to as non-IoT devices, which include laptops, smartphones, and tablets [5, 6]. With IoT devices and networks becoming the norm, due to the fast-declining prices of microchips and rapid advancements of different supporting technologies, new approaches and methods to provide more foolproof security solutions against these new threats are the need of the hour. Researchers [7] have highlighted that IoT and non-IoT devices will reach approximately 30.9 billion and 10.3 billion, respectively, by the year 2025. McKinsey and IHS Markit telecommunication industry have reported that IoT technologies will impact the global economy to the tune of around 11.1 trillion United States (US) dollar by the year 2025 [8], with about 125 billion IoT devices will be connected in the year 2030 [9]. These high projections suggest that IoT will become an integral part of everyday life in the near future.

This technological revolution has also led to an increase in the number of malicious activities and attacks on IoT networks. To effectively detect and mitigate these malicious activities, it is essential to have a robust and accurate method for identifying network-connected devices. Device fingerprinting (DFP) is a technique that may be used to identify and track IoT devices based on unique characteristics of the device, by utilizing the internet protocol (IP) address, media access control (MAC) address, operating system, network traffic features (packet, frame, and signal) and browser type [5, 10, 11, 12]. By processing and combining these pieces of information, a unique device fingerprint can be formulated that can be used to identify the device even if its explicit identification information, such as IP and MAC addresses, has been maliciously altered. This technique is often used for security and tracking purposes, including for the identification, and blocking of malicious devices.

In this paper, an efficient machine learning (ML) based device fingerprinting (DFP) model has been proposed for classifying individual IoT devices using only a single transmission control protocol (TCP)/IP packet header information. The proposed model uses only device-originated packet traces for generating unique fingerprints, with 22 features selected using a gain-ratio attribute evaluator and empirical anal-ysis of the datasets. Overall, the proposed ML-based DFP scheme obtained 83.70% and 98.20% accuracies on the IoT Sentinel [13] and UNSW [5] datasets, respectively. The proposed DFP model is undoubtedly able to assist in improving network security as well as able to be used for the management of a network. In brief, the key contributions of this research work are:

1. Identifying a suitable subset of network traffic features from a large number of features pool for generating unique fingerprints, using only a single device-originated TCP/IP packet header, to classify IoT devices,
2. Evaluation of the proposed DFP model using a tree-based supervised ML algorithm with two publicly available datasets, and
3. Evaluation of the features not only using mathematical operations but also empirical analysis with different datasets for identifying the significance of the individual feature.

The remainder of this paper is organized as follows. **Section 1.2** discusses related works in the area of device fingerprinting. The proposed machine learning-based IoT device identification model and datasets are given in **Section 1.3**, followed by discussions on the classification performances of the proposed model in **Section 1.4**. Finally, **Section 1.5** concludes the paper.

## 1.2　RELATED WORK

Device fingerprinting can be generated from information obtained from the different layers of the communication model, including network packets, MAC frames, and radio signals. Due to the wide availability and low-cost nature of hardware needed to capture network traffic traces, many researchers have proposed different DFP schemes using only network traffic traces.

Miettinen et al. [13] proposed an ML-based device fingerprinting model for classifying individual smart home IoT devices using network traffic features. 276-dimensional feature vectors (12 packets x 23 features) were utilized to generate unique fingerprints, with these fingerprints used to train an ML model per device type. The model reported 81.5% accuracy (global ratio) on the IoT Sentinel dataset with 27 IoT devices. Similarly, Aksoy and Gunes [14] proposed a DFP model based on the analysis of passively observed network packet traces, whereby 212 features were selected from a single TCP/IP packet's information for generating unique fingerprints. A J48 classifier was utilized for training and testing their proposed model. Overall, 82% accuracy was reported on the IoT Sentinel dataset with a set of 23 IoT devices.

On the other hand, Sivanathan et al. [15] presented a DFP model based on a statistical analysis of network traffic characteristics, including device sleeping time, packet size, and domain name system (DNS) intervals, for generating fingerprints. 12 statistical features were calculated from daily observed packet traces. The proposed DFP model reported over 95% accuracy using the UNSW dataset with 21 IoT devices. Similarly, Sivanathan et al. [5] presented a DFP approach also by utilizing 8 statistical features but calculated from hourly-based traffic traces. The method reported over 99% accuracy (UNSW dataset with 28 devices).

Table 1.1  Existing DFP approaches.

| Source | Objective | FSA | Fingerprint | Devices |
|---|---|---|---|---|
| [13] |  | Manually | $12^{Pkt} \times 23^{Feat}$ | $27^{IoT\ Sentinel}$ |
| [14] |  | Genetic Algorithm | $1^{Pkt} \times 212^{Feat}$ | $23^{IoT\ Sentinel}$ |
| [15] |  | Manually | $1Day^{(nPkt)} \times 12^{Feat}$ | $21^{UNSW}$ |
| [5] | Device | Manually | $1Hour^{(nPkt)} \times 8^{Feat}$ | $28^{UNSW}$ |
| [4] | Identification | Metric-entropy | $1^{Pkt} \times 161^{Feat}$ | $27^{IoT\ Sentinel}$ |
|  |  |  | $1^{Pkt} \times 86^{Feat}$ | $19^{UNSW}$ |
| [10] |  | Gain Ratio, Sorting | $1^{Pkt} \times 24^{Feat}$ | $31^{IoT\ Sentinel}$ |
| [16] |  | – | $100^{Pkt} \times 2^{Feat}$ | $21^{UNSW}$ |

Note: Feature Selection Algorithm - FSA, Packet - Pkt, Feature - Feat

In reference [4], an ML-based DFP model was proposed by selecting 161 features from a total of 212 features of a single TCP/IP packet, using the three metrics operations for generating fingerprints. Classification performance reached over 83% accuracy (J48 classifier) using the IoT Sentinel dataset with 27 IoT devices, and over 97% accuracy using the UNSW dataset with 19 IoT devices. On the other hand, Chowdhury et at. [10] presented an ML-based model but using only 24 packet-level features to train the ML model for the classification task. The model reported 83.9% accuracy (J48 classifier) using the IoT Sentinel dataset with 31 IoT Devices. The same researchers [16] subsequently proposed a deep learning-based model for the classification task, using only two network packet features: *tcp.window size* and *ip.len*, by utilizing 100 consecutive packets for generating unique fingerprints. Fingerprints were presented as graphs. The model gained over 98% classification performance using the UNSW dataset with 21 smart-home distinct IoT devices.

From the existing works, as presented in Table 1.1, it has been observed that the feature selection process for generating unique device-specific fingerprints is a challenging task and impacts the overall classification model performances. Researchers have utilized different methods, including mathematical models (such as metric entropy, and threshold value), genetic algorithm (GA), gain ratio, and sorting algorithms, for selecting a suitable subset of features to classify individual devices. These researchers have been analyzed a large set of features either from a single packet or *n* number of packets for generating unique fingerprints, whilst a large number of features not only increase computational complexity but decrease classification performances. In this study, a suitable subset of the most significant features has been selected using a gain ratio attribute evaluator and a ranking algorithm, whilst these features set (22 features) have been extracted only a single TCP/IP packet header information. Insignificant features have been subsequently removed empirically from the selected list, for generating unique fingerprints and have been shown to give respectable classification performances.

## 1.3 METHODOLOGY

The proposed DFP model uses passively observed network traffic traces for generating unique fingerprints, with these fingerprints/signatures used for training a supervised ML model and, subsequently, for testing the proposed DFP model on different publicly available datasets. ML approaches have been utilized for different purposes, including device classification [10, 12], network traffic classification [17], malicious traffic classification [18], and recommendation systems [19]. This section describes the datasets, the supervised classification algorithm, and the proposed DFP model.

### 1.3.1 IoT Datasets

Two publicly available datasets, IoT Sentinel [13] and UNSW [5] datasets were utilized to evaluate the proposed DFP model. A summary of these datasets is presented in Table 1.2. The IoT Sentinel dataset consists of 31 smart home IoT devices set up phase traffic traces, whilst the UNSW dataset comprises benign traffic traces from 22 IoT devices, including the TP-Link camera, smart bulb, Belkin camera, smart doorbell, printer, and smart photo frame.

Table 1.2  List of IoT datasets.

| Dataset | Number of Devices | Packets | Source |
|---|---|---|---|
| IoT Sentinel | 31 | 102,347 | [13] |
| UNSW | 22 | 6,845,378 | [5] |

### 1.3.2 Machine Learning Classifier: J48 (C4.5)

The Weka tool is used for the classification task. In Weka, the C4.5 classifier is implemented as J48, an extended version of the Iterative Dichotomiser 3 (ID3) algorithm [20]. It produces a decision tree for the classification task, with leaves and branches representing classification rules and features. The J48 classifier utilizes in-formation theory and tree pruning techniques to establish a decision tree for the classification tasks. Each attribute is selected based on information gain, with the classifier used to make the final decision [21]. Due to its simple implementation and high classification performance, the J48 has been utilized for different purposes, including device classification [4, 13], anomaly detection [22], and disease [23] classification tasks. Additionally, the J48 also allows the investigation of significant features (attributes) that may subsequently be utilized for the classification tasks.

### 1.3.3 Device Fingerprinting Model

A supervised ML-based device fingerprinting model architecture, depicted in Fig. 1.1, has been proposed. It uses device-originated traffic traces for generating unique fingerprints to identify individual devices. Device-originated traffic traces were collected passively using a network traffic analyzer tool, such as Wireshark, and then a total of 24 features [10] were extracted from a single TCP/IP packet using the TShark util-ity. Subsequently, these feature sets, as a vector of 24 features from a single packet, were labelled with the individual device name, with network packets grouped based on MAC addresses of the individual devices. The feature set was evaluated using the gain ratio attribute evaluator and ranked in descending order for identifying the most significant features. Two insignificant features: *tcp.options.timestamp.tsval* and *tcp.options.timestamp.tsecr*, were removed empirically (time-based features values) from the feature list for generating unique fingerprints. Naturally, this reduction reduces computational complexity. Subsequently, the selected 22 packet header features (a fingerprint) were used to train an ML model and used to evaluate the model performance on the IoT datasets. Table 1.3 lists the 22 selected features. Each dataset is randomly split into 80:20 ratios for training and testing to avoid bias or overfitting problems.

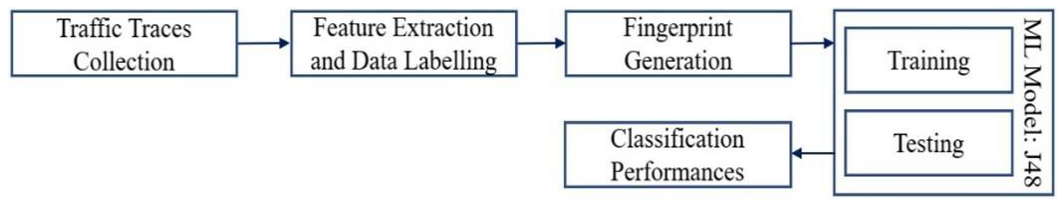

Figure 1.1  The proposed ML-based DFP model architecture.

## 1.4  RESULTS AND DISCUSSION

The performance of the proposed ML-based DFP model was evaluated on a Linux machine (Dual-core Intel i5-5200U CPU at 2.20GHz, DDR3L 16GB RAM, and 1TB hard drive), using a benchmark data mining tool, Waikato environment for knowledge analysis (WEKA) [20]. A tree-based classifier J48 was utilized to evaluate the classification performance. Two publicly available datasets, including IoT Sentinel and UNSW datasets, were utilized for evaluation. These datasets were divided into two subsets randomly, with 80% of the data utilized for training and the rest utilized for testing.

The performances of the proposed supervised ML-based DFP method in classifying individual IoT devices on the different datasets are depicted in Fig. 1.2. In Fig. 1.2, it is observed that the proposed DFP model achieved 83.9% accuracy in identifying individual IoT devices using a set of 22 features on the IoT Sentinel dataset (31 devices), whilst using a set of 24 features on the same classifier the reference model provided 0.2% higher accuracy. The low classification performances (around 84% accuracy) are due to the availability of similar types of devices on the IoT Sentinel dataset. Additionally, this dataset has a limited number of instances from the different IoT devices for training and testing. For instance, iKettle2 and SmarterCoffee devices have only 60 and 61 instances, respectively.

On the other hand, the proposed DFP model achieved over 1.2% higher accuracy than the reference model on the UNSW dataset with a set of 22 smart home

Table 1.3  List of network packet header features.

| Feature Name | Protocol | OSI Model |
|---|---|---|
| *http.request_number*<br>*http.prev_request_in* | HTTP | Application |
| *udp.srcport*<br>*udp.stream*<br>*udp.length*<br>*udp.dstport*<br>*udp.checksum* | UDP | Transport |
| *tcp.srcport*<br>*tcp.stream*<br>*tcp.dstport*<br>*tcp.window_size*<br>*tcp.ack*<br>*tcp.window_size_scalefactor*<br>*tcp.window_size_value* | TCP | Transport |
| *ip.len*<br>*ip.dsfield.dscp*<br>*ip.hdr_len*<br>*ip.dsfield*<br>*ip.id*<br>*ip.ttl*<br>*ip.proto*<br>*ip.dsfield.dscp* | IP | Network |

Note: Open Systems Interconnection - OSI

IoT devices. It is worth noting that the UNSW dataset consists of different types of IoT devices from different manufacturers with a large number of instances from individual devices. Consequently, the proposed model shows significant classification performances in classifying individual IoT devices, despite the dataset consisting of different types of devices.

The classification performance of the proposed DFP model was also evaluated using a rules-based classifier, i.e. Decision Table. Both publicly available datasets, including IoT Sentinel and UNSW IoT datasets, are utilized for evaluating this model. In Fig. 1.3, it has been observed that the proposed DFP model achieved over 89% and 96% accuracies on the IoT Sentinel and UNSW datasets, respectively. This classifier shows higher accuracy compared to the J48 classifier on the IoT Sentinel dataset, but devices with a limited number of instances, such as iKettle2 and Smarter Coffee IoT devices, cannot be identified accurately. Conversely, the proposed model achieved almost similar results on the UNSW dataset. From empirical and literature studies, it is worth noting that the IoT Sentinel dataset is a challenging dataset in terms of data limitation and available instances from different IoT devices (network packets), which consists of only setup phase traffic traces. On the other hand, the UNSW dataset consists of a large number of instances from different devices, hence the proposed DFP model can learn communicate behaviour with high accuracy.

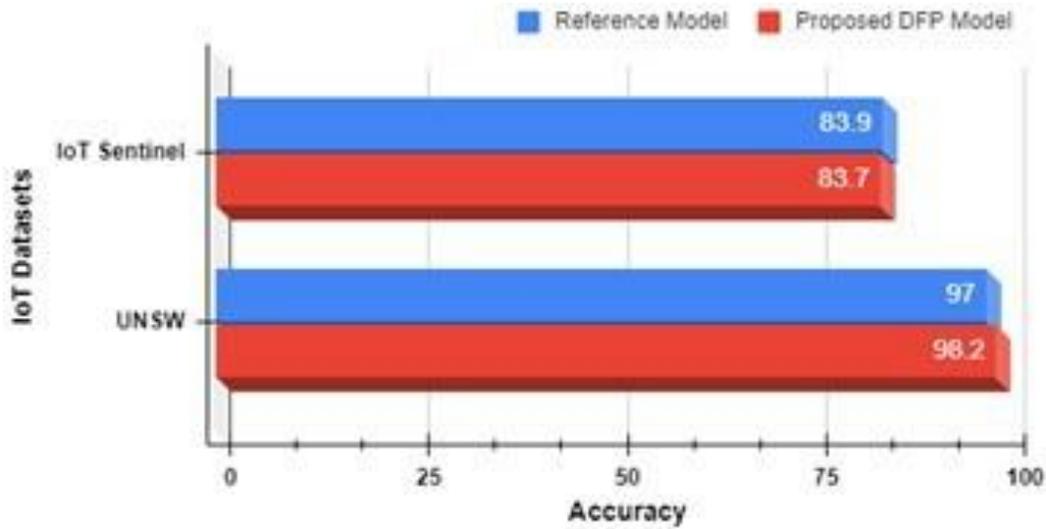

Figure 1.2  Classification performance (J48 Classifier) of the IoT devices: IoT Sentinel and UNSW datasets.

Table 1.4  Comparison of the device fingerprinting models.

| Source | Fingerprint | Devices/Datasets | Performance |
|---|---|---|---|
| [13] | $12^{Pkt} \times 23^{Feat}$ | $27^{IoT\ Sentinel}$ | $81.5\%^{IoT\ Sentinel}$ |
| [14] | $1^{Pkt} \times 212^{Feat}$ | $23^{IoT\ Sentinel}$ | $82\%^{IoT\ Sentinel}$ |
| [15] | $1Day^{nPkt} \times 12^{Feat}$ | $21^{UNSW}$ | $95\%^{UNSW}$ |
| [5] | $1Hour^{nPkt} \times 8^{Feat}$ | $28^{UNSW}$ | $99\%^{UNSW}$ |
| [4] | $1^{Pkt} \times 161^{Feat}$ | $27^{IoT\ Sentinel}$ | $83.35\%^{IoT\ Sentinel}$ |
|  | $1^{Pkt} \times 86^{Feat}$ | $19^{UNSW}$ | $97.78\%^{UNSW}$ |
| [10] | $1^{Pkt} \times 24^{Feat}$ | $31^{IoT\ Sentinel}$ | $83.9\%^{IoT\ Sentinel}$ |
| [16] | $100^{Pkt} \times 2^{Feat}$ | $21^{UNSW}$ | $98.54\%^{UNSW}$ |
| Proposed DFP Model | $1^{Pkt} \times 22^{Feat}$ | $31^{IoT\ Sentinel}$ $22^{UNSW}$ | $89.61\%^{IoT\ Sentinel}$ $96.33\%^{UNSW}$ |

Note: Packet - Pkt, Feature - Feat

Table 1.4 presents a comparative summary of the existing DFP approaches along with the proposed ML-based DFP model. Prior approaches have utilized a large number of features either extracted from a single packet or n-number of packet information for generating unique device fingerprints. It can be seen that the proposed DFP model shows a higher accuracy of 89% using the IoT Sentinel dataset compared to other models presented in references [13, 14, 4, 10], whilst using only 22 selected features from a single TCP/IP packet. On the other hand, the proposed model also provides over 96% accuracy using the UNSW dataset, though in references [15] and

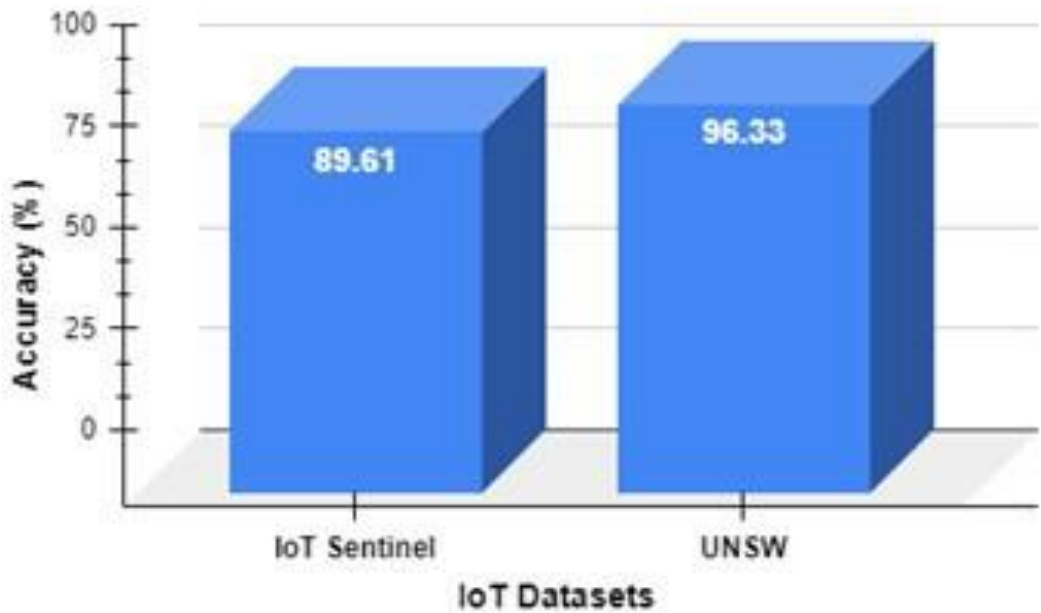

Figure 1.3 Classification performance (Decision Table) of the IoT devices: IoT Sentinel and UNSW datasets.

[5] the models gain 2.21% and 2.67% higher accuracies with a large number of packet information.

## 1.5　CONCLUSION

In a network, a large number of heterogeneous IoT devices are being connected to obtain network-based services. However, these interconnections impose different types of security challenges for network administrators and operators, including the identification of malicious devices. Traditional identifiers, such as IP and MAC addresses, can be easily mutated and hence, are not a defense against these malicious devices. Some researchers have proposed network traffic-based features engineering for generating unique fingerprints to identify IoT devices. However, it is challenging to identify device-specific features from a large pool of features. In this paper, an ML-based DFP model has been proposed to identify network-connected IoT devices using only network traffic characteristics from a single TCP/IP packet header. This selection of a subset of device-specific features reduces computational complexity but has also been shown to improve classification performances. A total of 22 features have been selected for generating unique fingerprints. 83.70% and 98.20% classification accuracies were reported on the IoT Sentinel and UNSW datasets, respectively. These respectable accuracies signify that the proposed model is useful for device identification and may be used by the network operator to improve the security of a network, especially against malicious devices.

**Acknowledgement**
The authors are profoundly grateful to the Department of Computer Science and Engineering (CSE), Metropolitan University (MU), Sylhet, Bangladesh, and the Faculty of Integrated Technologies (FIT), Universiti Brunei Darussalam (UBD), Gadong, Brunei Darussalam, for supporting this research work.

**Conflict of Interest**
There are no conflicts of interest from the authors.